# Reply to "Comment on "Extension of the Bethe-Weizsäcker mass formula to light nuclei and some new shell closures"

C. Samanta[1,2] and S. Adhikari[1]

[1]*Saha Institute of Nuclear Physics, 1/AF Bidhan Nagar, Kolkata – 700 064, India*
[2]*Physics Department, Virginia Commonwealth University, Richmond, VA 23284, USA*

Some properties of the modified Bethe-Weizsäcker mass formula (BWM) are discussed. As BWM has no shell effect included, the extra-stability or, magicity in nuclei clearly stands out when experimental mass data are compared with BWM predictions. If the shell effect quenches, the BWM predictions come closer to the experimental data.

The paper by C. Samanta and S. Adhikari [1] begins by highlighting the inadequacy of the Bethe-Weizsäcker mass formula (BW) for light nuclei and extends the BW for light nuclei by modifying its symmetry and the pairing terms. The beauty of this new mass formula is that by introducing only two parameters, the formula improves drastically compared to the Bethe-Weizsäcker mass formula. The modified Bethe-Weizsäcker mass formula (BWM) reproduces the trend of the binding energy versus neutron number curves of almost all the nuclei from Li to Bi. It is clearly stated in Ref. [1] that as the shell effect is not incorporated in BWM, it does not reproduce the binding energies of the magic nuclei.

The stability of a nucleus near the drip line is assessed by the evaluation of one- or, two-nucleon separation energies, i.e., $S_p$ ($S_n$) and $S_{2p}$ ($S_{2n}$). The BWM results deviate from the experimental values if any nucleus involved in this evaluation has large shell effect or, magicity. The disagreement of $^{5}$Li, $^{6}$Be stability with the experiment occurs due to N=2 magicity of $^{5}$Li, $^{6}$Be. The $S_p$ or, $S_{2p}$ calculations of $^{15}$F and $^{16}$Ne deviate from the experimental values as the said calculations need the binding energy of the magic nucleus $^{14}$O [1]. The experimental binding energies of $^{19}$B, $^{22}$C are larger compared to BWM due to increasing shell effects in these nuclei approaching the N=14 and 16 magicities. The cases of $^{31}$F and $^{26}$O are discussed in Ref. [1].

It must be pointed out that as there is no shell effect in BWM, it is not at all meaningful to compare the chi-square of the BWM and a macroscopic-microscopic model which has the shell effect incorporated. The chi-square ($\chi^2$) of macroscopic formulas like improved liquid drop model (ILDM) [2] and BWM are expected to be larger. However, the chi-square of BWM ($\chi^2_{BWM}$ = 12.94) is significantly less than the chi-square of ILDM ($\chi^2_{ILDM}$ = 29.46). (In our subsequent calculations [3] we used the optimized parameter $a_v$ =15.777 MeV instead of 15.79 MeV for which the $\chi^2_{BWM}$ reduces to 11.10). Compared to

ILDM [2], the BWM yields a better fit to the binding energy versus nucleon number curves for light nuclei. Thus the BWM can be used in calculations where only a macroscopic formula is needed over a wide mass range. When the shell effect quenches, the BWM predictions come closer to the experimental masses. Thus BWM is also useful to determine the "shell effect quenching" in nuclei [3].

In BWM, the binding energy is defined as,

$$BE(A,Z) = 15.777A - 18.34A^{2/3} - 0.71\frac{Z(Z-1)}{A^{1/3}} - 23.21\frac{(A-2Z)^2}{A(1+e^{-A/17})} + \delta_{new}$$

where, $\delta_{new}$ = +12 $A^{-1/2}$(1-$e^{-A/30}$) for even Z-even N nuclei, $\delta_{new}$ = -12 $A^{-1/2}$(1-$e^{-A/30}$) for odd Z-odd N nuclei and, 0 for odd A nuclei. It is well known to the physics community that a mass formula based on the liquid-drop (LD) model is not expected to explain the finer details. Nevertheless, it is surprising to note how well a simple LD model-type formula remains applicable even far away from the stability line. For example, the one-proton separation energy versus proton number curve is remarkably well reproduced by BWM, around Z=82 for N=106 and 107 [4] and, the experimental binding energies of $^{188}$Pb (1470.824 MeV), $^{189}$Pb (1479.095 MeV), $^{187}$Tl (1468.135 MeV), $^{188}$Tl (1476.406 MeV) are reproduced by BWM within 56 keV, 257 keV, 305 keV and 347 keV respectively, as the shell effect in these nuclei are greatly reduced [3]. In the recently published 2003 edition of the nuclear mass tables [5] the above numbers have changed to $^{188}$Pb (1471.070 MeV), $^{189}$Pb (1479.206 MeV), $^{187}$Tl (1468.410 MeV), $^{188}$Tl (1476.388 MeV) and these experimental binding energies are reproduced by BWM within 302 keV, 146 keV, 31 keV and 366 keV respectively.

References:

1. C. Samanta and S. Adhikari, Phys. Rev. **C 65**, 037301 (2002)
2. S. R. Souza et al., Phys. Rev. **C 67**, 051602(R) (2003)
3. C. Samanta and S. Adhikari, nucl-th/0402016; Nucl. Phys. A (in press)
4. C. Samanta, Acta. Phy. Hung. **19**, 161 (2004); ibid, Proceedings of the Symposium on Nuclear Clusters: From Light Exotic to Superheavy Nuclei, Rauischholzhausen, Germany, 5-9 August, 2002, page 467, Ed. Jolos and Scheid, (ISBN 963 206 299 X)
[5] G. Audi, O. Bersillon, J. Blachot and A. H. Wapstra, Nucl. Phys. A729 3 (2003)